# Fractal String Generation and Its Application in Music Composition

*Avishek Ghosh, Joydeep Banerjee, Sk. S. Hassan, P. Pal Choudhury*

*Applied Statistics Unit*

*Indian Statistical Institute, Calcutta*

## *Abstract*

Music is a string of some of the *notes* out of 12 *notes* (Sa, Komal_re, Re, Komal_ga, Ga, Ma, Kari_ma, Pa, Komal_dha, Dha, Komal_ni, Ni) and their *harmonics*. Each note corresponds to a particular *frequency*. When such strings are encoded to form discrete sequences, different frequencies present in the music corresponds to different *amplitude* levels (value) of the discrete sequence. Initially, a class of discrete sequences has been generated using logistic map. All these discrete sequences have at most n-different amplitude levels (value) (depending on the particular raga). Without loss of generality, we have chosen two discrete sequences of two types of Indian raga viz. Bhairabi and Bhupali having same number of amplitude levels to obtain/search close relatives from the class. The relative / closeness can be assured through correlation coefficient.The search is unbiased, random and non-adaptive. The obtained string is that which maximally resembles the given two sequences. The same can be thought of as a music composition of the given two strings. It is to be noted that all these string are fractal string which can be persuaded by fractal dimension.

## *1. Introduction and Review:*

Music is a field of study that has an obvious relationship to mathematics. Music is, to many people, a nonverbal form of communication, which reaches past the human intellect directly into the soul. However, music is not really created by mankind, but only discovered, manipulated and reorganized by mankind. In reality, music is first and foremost a phenomenon of nature, a result of the principles of physics and mathematics. A musical scale is a discrete set of pitches used in making or describing music.Precisely, Music can be encoded as a string of at most 12 notes viz. (Sa, Komal_re, Re, Komal_ga, Ga, Ma, Kari_ma, Pa, Komal_dha, Dha, Komal_ni, Ni) and their harmonics [1, 2, 3, 4]. In this paper, without loss of generality, two different types of Indian raga have been considered. Initially, two music strings of 1000 length have been taken from each of the raga. A class of discrete sequences has been generated through an illustrious chaotic map viz. logistic map. From the class, we have devised a mechanism to find out a close relative to the two those strings. The mechanism is based on correlation coefficient. Before going into the technical details, let us revisit logistic map very briefly.

**Logistic Map:**The one dimensional, polynomial map of degree 2 is given as

$$l_{n+1} = \lambda . l_n . (1 - l_n)\ n \geq 0$$

Where $l_n$ belongs to the (0, 1) and $l_0$ represents the initial value. $\lambda$ is a positive number, and represents as a control parameter of the chaotic system. The dynamical behaviour depends on entirely on the value of $\lambda$ [5, 6].

## *1.1 String generation using Logistic Map*:

Using Logistic map a set of sequence of length 1000 have been generated. Then music strings corresponding to different ragas can be generated. The work is carried on withtwo famous Indian raga viz. *Bhairabi and Bhupali*.

### *1.1.1 Bhairabi:*

In Bhairabi raga, there are seven fundamental frequencies. They are *Sa, Komal_re, Komal_ga, Ma, Pa, Komal_dha, and Komal_ni*. So a music string in Bhairabi raga consists of these seven notations along with the Octet Sa (Sa') and Pause (corresponds to zero frequency). It is evident that in the Logistic map an one dimensional string of number between (0, 1) would come out. Accordingly, Logistic map with 9 frequencies and 1000 length is used to generate various music strings in Bhairabi raga. In the strings Sa, Komal_re, Komal_ga, Ma, Pa, Komal_dha, Komal_ni, Octet Sa and Pause are encrypted as S, r, g, M, P, d, n, C and B. The frequencies allotted for different encrypted notes in Bhairabi are given in Table-1.

| ***Notes (Encrypted)*** | ***Threshold*** |
|---|---|
| S | (0, 1/9) |
| R | (1/9, 2/9) |
| G | (2/9, 1/3) |
| M | (1/3, 4/9) |
| P | (4/9, 5/9) |
| D | (5/9-2/3) |
| N | (2/3, 7/9) |
| C | (7/9, 8/9) |
| B | (8/9, 1) |

**Table-1:** The frequencies allotted for different encrypted notes in Bhairabi

Examples of such two strings are given:

*Bhairabi_1 (corresponding to* $\lambda = 3.99, initial\ point = 0.1$):

```
SMBgCPBSrPBSSgCMBrMBSgCdBrMBSgCPBSrdBMBrdBMBMBgCMBSgCPBSrPBSSrPBSSrdBr
MBrPBSrdBSMBrdBgCMBrPBSSrPBSSrdBMBgnCdBrdBrMBSgCPBSSMBgCPBSSgCPBSrMBSM
BMBgCMBSgnnCdBrdBgCPBSSrPBSrMBSgnnCdBSMCMBrPBSrPBSSgCdBrnCPBSSrdBSgnnn
nCdBrPBSrPBSrdBrMBSgCnCPBSSgCdBMBrdBMBgCdBrPBSrMBSgCnCMBrPBSSrdBrdBSMB
gnnnnCdBrdBSMBgnCnCPBSrPBSSgnCdBgnnnnnCdBgCnCPBSSrdBSgCdBrdBrdBSMBgnCd
BMBgCPBSSMBMBgCdBgnCdBMBMBgnCdBgCdBgCPBSrPBSSMBgnnCdBgCMBSgCPBSrMBrdBS
gCPBSrMBrPBSrnCPBSrPBSSgCdBrnCMBSMBgCMBrMBrPBSSrPBSSrdBrMBrdBSgCnCPBSS
rdBrMBSgnCdBSgnnCPBSrMBrPBSrMBSMBMBMBMBrdBrPBSSMBgnnCPBSrdBgCdBrPBSrMB
rPBSrPBSSrPBSrPBSSgCMBSgnnCPBSSrdBrPBSSMBrnCPBSrPBSrPBSrdBgnnCPBSSgCMB
SgnCdBrPBSrdBgnCdBgCdBgCdBgCdBrdBMCMBrdBrMBrPBSSrPBSrMBSgnnnnnnCdBgCPB
SSgCnCPBSrdBMBgCMBSgCPBSSgnCdCMBrdBSgCPBSrMBSgCdBgCdBgnnnCdBrdBSMBrdBr
PBSSgCPBSSrPBSrdCMBrPBSrPBSrPBSSgnnCnCPBSrdBgCMBrPBSrdBrPBSrPBSrPBSSMB
rdBgCPBSrdBgnCdBgCdBrdBSgnnnnCPBSSrdBgnCPBSSrPBSSMBrdBgCPBSSrdBrMBrMBr
PBSrPBSSgCPBSSrPBSSrdBgCPBSrMBSgCdBgnCdBSgnCPBSSMBgnCdBrdBgCMBSMBgnnnC
PBSrdBSMBMBgnCPBSrPB
```

*Bhairabi_2(corresponding to* $\lambda = 3.95, initial\ point = 0.3$) *:*

```
gCdBSMBMCMBrPBSrdBgCnCPBSgnCdBgCnCPBSgCdBrPBSrdBMBgCdBMBgnCdBrPBSgCdCM
BrPBSgCMBSMBgnnnCdBgCPBSgnCdBgCPBSgCdBgCPBSgnnnnnCdBrPBSrdBrdBrMBrdBgC
dBgCPBSgnCPBSgnCnCPBSrdBMBgnnCPBSrdBgnnnnCdBSMBgCPBSgnCdCMBrdBgnnnCPBS
gCdBgCPBSgnnnCdBrPBSgCPBSgnnnnnnCdBSMBgCdBrPBSgnCPBSgCPBSrdBrdBrnCMBrM
BrMBrdBgCdBMBgnnnCPBSgnCdBrdBgCdCMBrPBSrdBrdCMBrPBSgnnnCdBrPBSgCdBMBgn
CPBSgnnCdBgnCPBSgCdBgnCdBgCdBrPBSrdBgCnCMBrMBSMBMBgCPBSgnnnCPBSrdBgCPB
SrdBrdBrdBrdBrPBSrdBgCdBgCdBMBMBMBgCPBSrdBgCdBrMBSMBMBMBgCdBSMBMBMCMBr
dBSMBMBgCPBSrdBrdBrPBSrdBrdBrMBrPBSgnnCdBgCPBSgCMBSMBgCnCPBSgnCPBSgCPB
SrdBgnCdBMBgnnnnCdBgCdBrdBgCdBrMBrnCPBSrdBgnnnnnnnnnCdBMBMCMBrPBSrdBr
dBSMBgCPBSgnnnCPBSMCMBrPBSrdBgnnnCPBSgCdBrPBSgCdBgCdBgCPBSrdBMBgCdBgnn
CPBSrnCPBSrdBMBMBgnnnCdBrdBrPBSrdBrdBrdBgnCdBSMBgCMBrMBrdBrdBMBMBgCdBS
MBgnCdBrdBgCPBSrdCMBrdBMBMBgnCdBrMBrdBSMBgCPBSgCMBrPBSgnnCdBgnnnnCdBgC
PBSrdBgnCnCMBrMBrMBrPBSrdBMBgCPBSrdBgnCPBSrdBrdBSMBgCdBgCnCMBSMBgnCdBr
PBSgCPBSrdBgCdBrMBrPBSgCnCPBSrdBMBMBMBMBMBgCPBSrdBgCPBSrdCMBrdBSMBMBgC
dBgCdBMBMBMBgCPBSgCP
```

The graphical representations of the logistic maps corresponding to the given strings are shown:

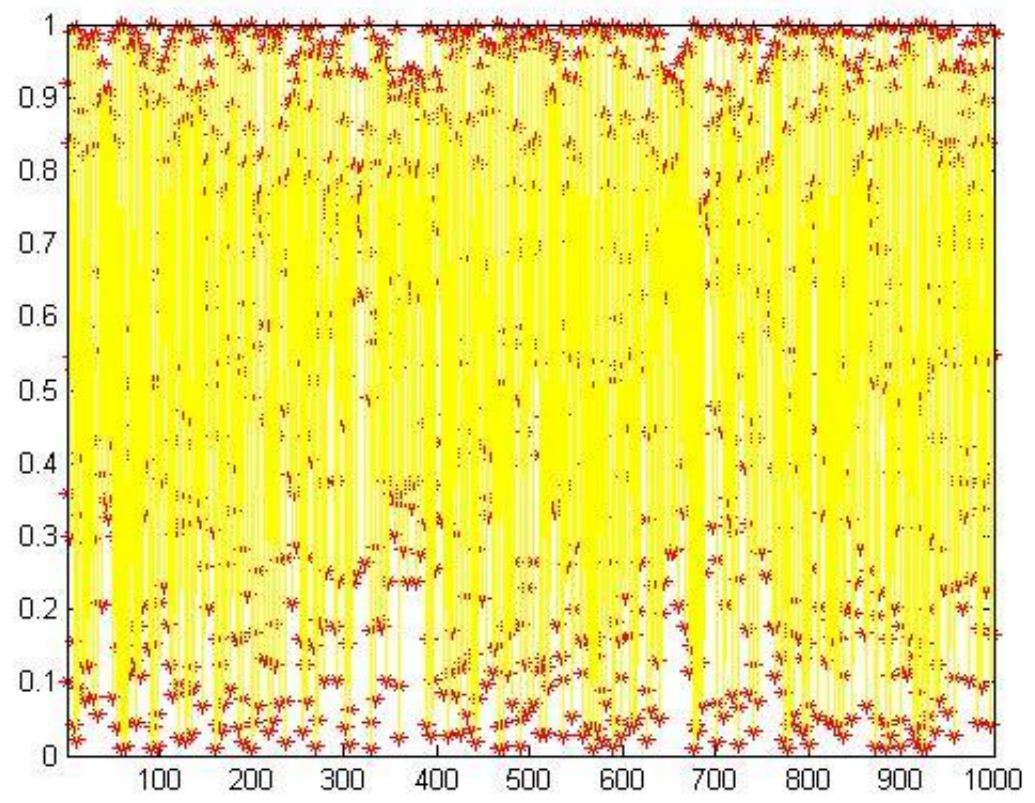

Bhairabi_1

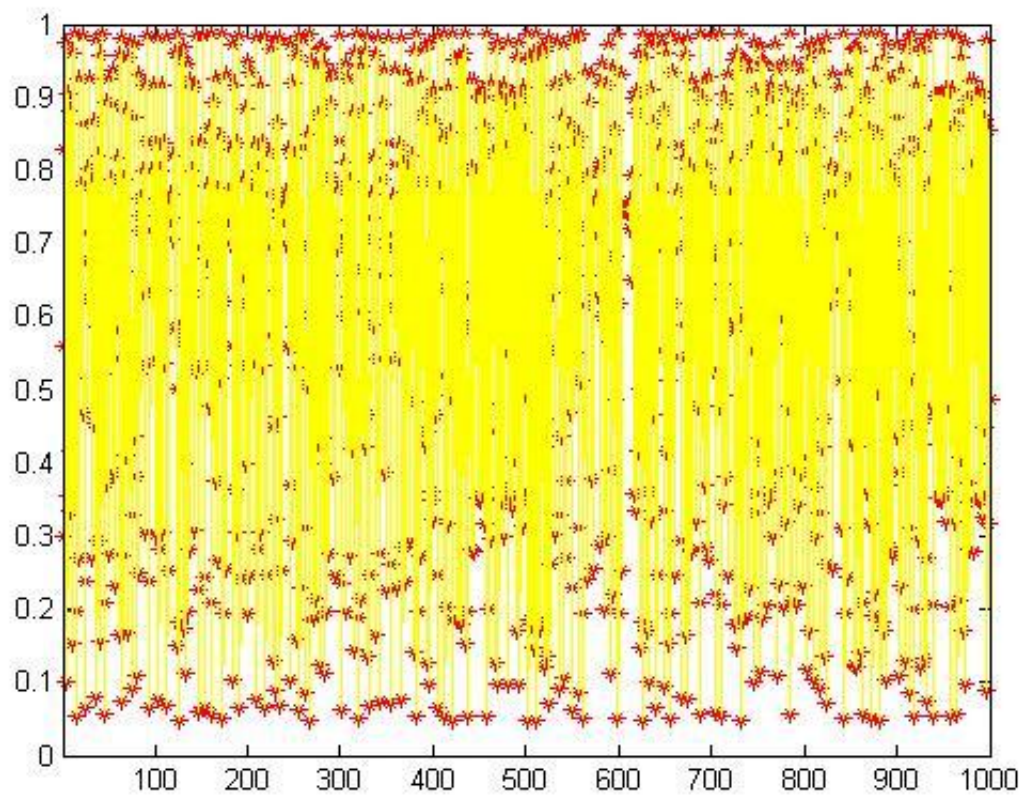

Bhairabi_2

### *1.1.2 Bhupali:*

In Bhupali raga, there are five fundamental frequencies. They are Sa, Re, Ga, Paand Dha. So a music string in Bhupali raga consists of these five notations along with the Octet Sa (Sa') and Pause (corresponds to zero frequency). Logistic map with 7 frequencies and1000 length is used to generate various strings of this particular raga. In the strings Sa, Re, Ga, Pa and Dha, Octet Sa and Pause are encrypted as S, R, G, P, D, C and B. The frequencies allotted for different encrypted notes in Bhupali are given in Table-2.

| ***Notes (Encrypted)*** | ***Threshold*** |
|---|---|
| S | (0, 1/7) |
| R | (1/7, 2/7) |
| G | (2/7, 3/7) |
| P | (3/7, 4/7) |
| D | (4/7, 5/7) |
| C | (5/7-6/7) |
| B | (6/7, 1) |

**Table-2.** The frequencies allotted for different encrypted notes in Bhupali

Out of them two strings are given below:

*Bhupali_1(corresponding to* $\lambda = 3.99, initial\ point = 0.1)$*:*

SGBGCPBSRPBSSGBGBSPBSGCDBSPBSGCPBSRDBGBRDBGBGBGBPBSGCPBSRPBSSRPBSSRDBS
GBSPBSRDBSGBRDBGBGBRPBSSRPBSSRDBGBRCCDBRDBSPBSGCPBSSGBGBPBSSGBPBSSGBSG
BGBGBPBSRCCCDBRDBGCPBSSRPBSSPBSRCDCDBSGBGBRPBSRPBSSRCDBRDCPBSSRPBSRCCC
DCDBSPBSSPBSRDBSPBSRCDBPBSSRCDBGBRDBGBGCDBRPBSSPBSRCDBGBRPBSSRDBRDBSGB
RCCCCCDBRDBSGBRCCDBPBSRPBSSRCCDBRCCCCCCDBRCDCPBSSRPBSRCDBRDBRDBSGBRCCD
BGBGCPBSSGBGBRCDBRCCDBGBGBRCCDBRCDBGBPBSRPBSSGBRCCCDBGBPBSGBPBSSGBRPBS
GCPBSSGBSPBSRDBPBSSPBSSRCDBRDBGBSGBGBGBSGBRPBSSRPBSSRDBSGBRPBSRCDCPBSS
RDBSPBSRDCPBSRCDCPBSSGBSPBSSGBSGBGBGBGBRDBSPBSSGBRCDCPBSRDBGCDBRPBSSGB
RPBSRPBSSRPBSSPBSSGBPBSRCDCPBSSRDBRPBSSGBRDBPBSRPBSRPBSRDBRCDCPBSSGBPB
SRDCDBSPBSRDBRCCDBRCDBRCDBRCDBRDBGBGBRDBSGBRPBSSRPBSSPBSRCCCCCCCDBGBPB
SSRCDBPBSRDBGBGBPBSGCPBSSRCCDBGBRPBSGBPBSSPBSRCDBRCDBRCCCCDBRDBSGBRDBS
PBSSGCPBSSRPBSRDBGBRPBSSPBSRPBSSRCCCDBPBSRDBGBGBSPBSRDBRPBSRPBSRPBSSGB
RDBGBPBSRDBRCCDBGCDBRPBSRCCCDCPBSSRDBRDCPBSSRPBSSGBRDBGCPBSSRDBSGBSGBR
PBSSPBSSGCPBSSRPBSSRDBGCPBSSPBSRCDBRDCPBSRDCPBSSGBRCCDBRDBGBGBSGBRCCDC
PBSRDBSGBGBRDCPBSRPB

*Bhupali_2(corresponding to* $\lambda = 3.95, initial\ point = 0.3)$:

GCPBSGBGBGBRPBSRDBRCDBPBSRCCDBRCDBPBSGCDBRPBSRDBGBRCDBGBRDCDBRPBSRCDBG
BRPBSGBPBSGBRCCCCDBGCPBSRCCDBGCPBSRCDBGCPBSRCCCCDCDBRPBSRDBRPBSGBRDBRC
DBGCPBSRDCPBSRCCDBPBSRDBGBRCDCPBSRDBRCCCDCPBSGBGCPBSRCCDBGBRDBRCCDCPBS
RCDBGCPBSRCCDCDBSPBSGBPBSRCCCCCDCPBSGBGCDBRPBSRDCPBSGCPBSRDBRDBRDBGBSP
BSGBRDBRCDBGBRCCDCPBSRDCDBRDBRCDBGBRPBSRDBRDBGBSPBSRCCDCDBRPBSRCDBGBRD
CPBSRCDCDBRDCPBSGCDBRCCDBGCDBRPBSRDBRCDBGBSPBSGBGBGBPBSRCCDCPBSRDBGCPB
SRDBRDBRDBRDBRPBSRDBRCDBRCDBGBGBGBGBPBSRDBGCDBSPBSGBGBGBGCPBSGBGBGBGBR
PBSGBGBGCPBSRDBRDBRPBSRDBRPBSGBSPBSRCCCDBGCPBSGBPBSGBRCDBPBSRDCPBSGCPB
SRDBRCCDBGBRCCCCCDBGCDBRDBGCPBSGBRDCPBSRDBRCCCCCCCCCCDBGBGBGBRPBSRDBR
PBSGBGBPBSRCCDCPBSGBGBRPBSRDBRCCDCPBSGCDBRPBSRCDBRCDBGBPBSRDBGBRCDBRCD
CPBSRDBPBSRDBGBGBRCCDCDBRDBRPBSRDBRDBRDBRDCPBSGBGBPBSGBRDBRDBGBGBGCPBS
GBRDCDBRDBGBPBSRDBGBRDBGBGBRDCPBSGBRPBSGBGCPBSGBGBSPBSRCCCDBRCCCCCDBGC
PBSRDBRCCDBPBSPBSGBSPBSRDBGBGCPBSRDBRDCPBSRDBRPBSGBRCDBRCDBPBSGBRDCDBS
PBSGCPBSRDBGCPBSGBSPBSRCDBPBSRDBGBGBGBGBGBGBPBSRDBGBPBSRDBGBRPBSGBGBRC
DBRCDBGBGBGBGBPBSGCP

The graphical representation of the logistic maps corresponding to the given strings is shown:

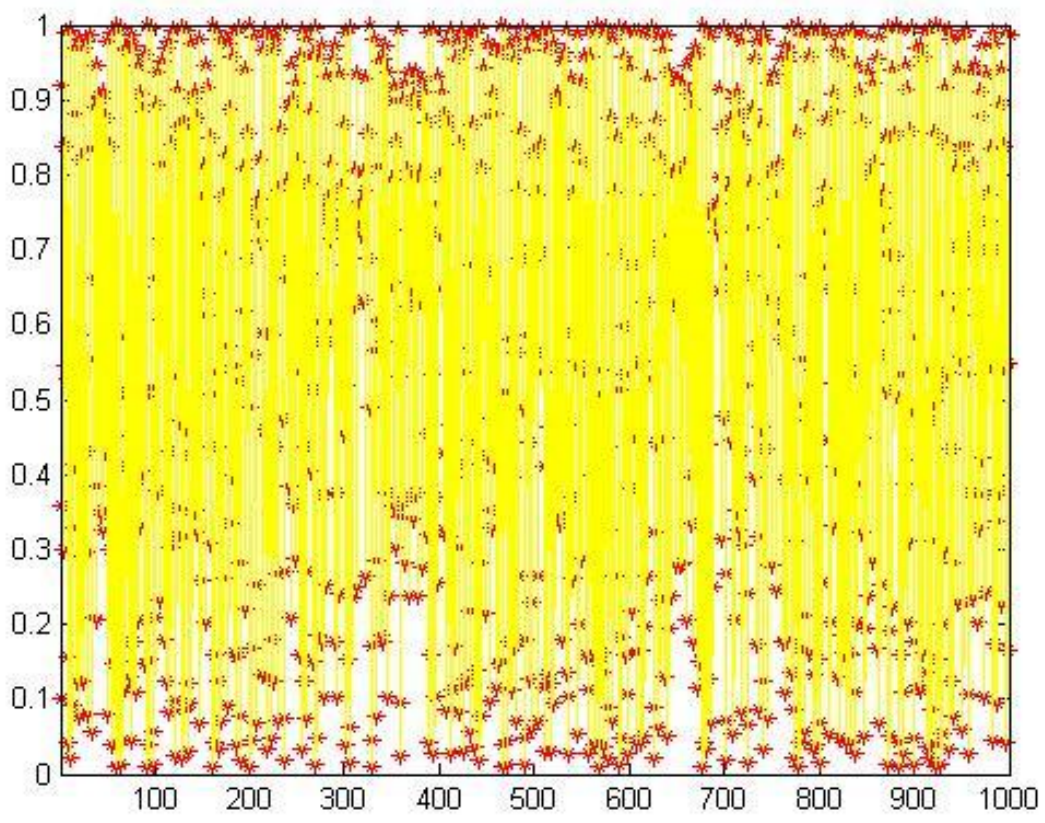

Bhupali_1

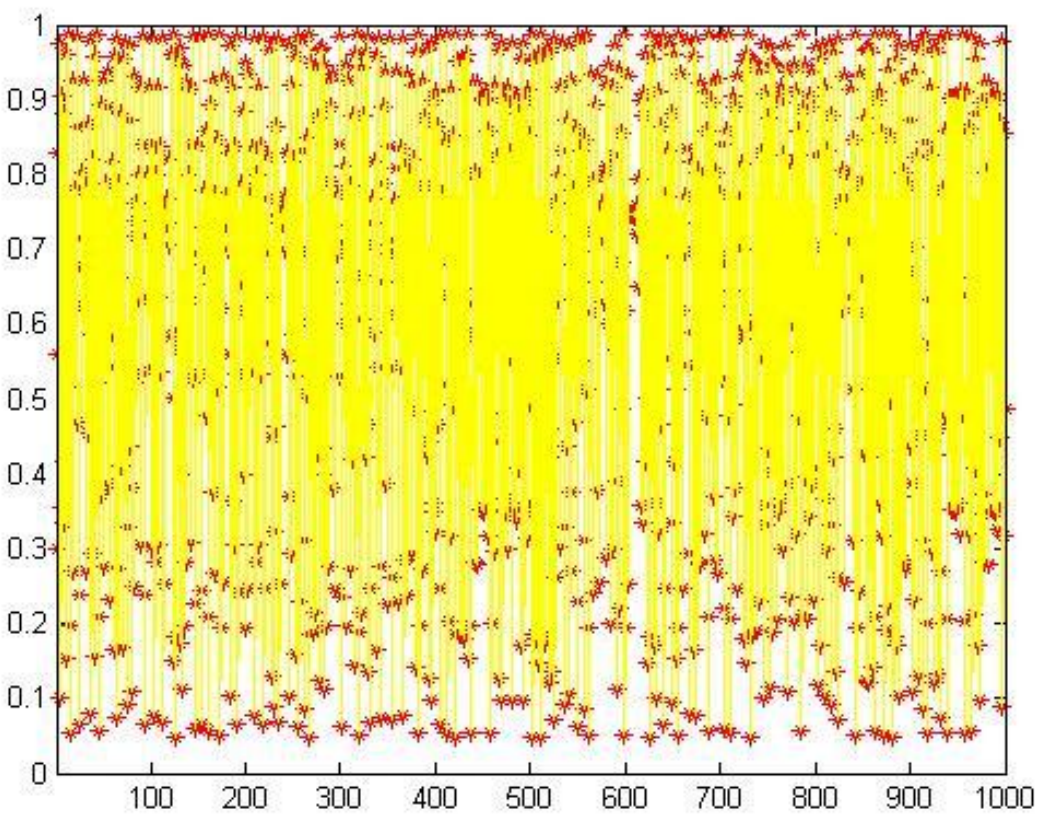

Bhupali_2

Now let us try to find out close relative string of the above two set of strings.

### *1.2 Obtaining a Close Relative String:*

The next task is to generate or evolve a new sequence from the given two sequences of the same class (same raga in present case). As said earlier the closeness is ensured by the parameter 'Correlation Coefficient'.

***Definition:*** Correlation Coefficient of two sequences $x_1(n)\ and x_2(n)$ is defined as

$$c = \frac{\sum_{n=1}^{1000} x_1(n)x_2(n)}{\sqrt{E_x E_y}}$$

Where $E_x$ and $E_y$ respectively denotes energy of $x_1(n)$ and $x_2(n)$ respectively, i.e.

$$E_x = \sum_{n=1}^{1000} {x_1}^2\,(n), and\ E_y = \sum_{n=1}^{1000} {x_2}^2\,(n)$$

Initially, the music strings are encoded to form discrete sequences where different frequencies present in the strings denote different amplitude levels of the sequence.Without loss of generality, we have chosen two discrete sequences having same number of amplitude levels (corresponding to music string of same raga) to obtain a close relative from the class.

#### *1.2.1 The Basic Algorithm:*

To get an evolved sequence from given two sequences of a given class, firstly a search space is randomly generated using typically 150000-200000 sequences of same class and same length (1000 in this case).The algorithm aids to find out the evolved or next generation sequence that maximally resembles the given 2 sequences, which can be assured from the fact that the sum of correlation coefficients of the evolved sequence with the given two sequences is maximum.The search is random, unbiased and non adaptive as the search space is randomly generated.According to the encoded logic, the derived sequence is finally decoded to a music string (with different frequencies present)of the same raga.

## 2. ***Results*:**

*Bhairabi*:

In case of Bhairabi raga, the close relative string corresponding to *Bhairabi_1* and *Bhairabi_2* is:

```
gBPBMPMgPPBMSCdnnMnrdrddSMgnSCCgCnrCBBrSMdrMMBCCBMnPCMdndSMSnMBCBCPBBM
nCSMnBgnnCnBgSPdBPrdCnMSgCSrdMgPSBSBrgMSCCnngndnCngMCSrMgdSMCPrnPCrBSr
nnrSPMgMSnMBSPPdrgBrCrBnSrrBMMdBdrnrgdgddrPPdrBCPMBMSSSnnPBdMMrrMMSBgn
rCBgdBMSrrdPBMMgrPgrSgngrCBrMCCrndSrSdMgPdggngrCrBrrgSCSrCSCSCgMPCPSgr
PdPPnBgSgCMgdrCPMSdgPBdgSSnCPggMggdngCBrSddnBPMSBnnrBCPMCddngnMgPrrPCn
MdgMgnBPPgPCMPCrPMMgdnrgCdrrSBCgnrrgCnSgrMSSnPMSMnPMCddgCPndBPPrMBBnrC
CSgdSBBrnMSBMPBSdPdCBdCPgBnPCddPBdnPndSggdrrCPddBdrnrrSPPSBCnrMgPngBCr
SnSnnPrMdrdrSnPSndnMMCdMSBMBgPSrBPBgCddrngnMnrrBgrPdCdrgnMndrBBMBnndCB
MSCBCMBrMMrdddMrSBnnnCPgdBMBBrrMrMBdgdnSPCndMgCdnnSPBBCnMrgBPCgMPPrPCM
rnrCgCMBPMBgMPCPdgdnrMSPdCBdBMPBdrBgnMnrBrgPSMSBrndgPddMPrPMCrnBnBndCM
CgBBSggBdgdgPrgSBgggBCMSSCPnPSPSPCBrrBdCrMCMPMrSBrrndnPdrPBCrgrBPMrCCS
SBnBnnCCnrrCBngSdBBSnrMgrBSrgCCPdBngPMgPSSBnCgCBPCCddMnndndnndBPMSdBrM
CPrBrPddMCPCnMddngBrCgCrnBPPnBnCrMPnPdnSMCSgMgMdnddgdPPgPBMgSgnMSPBBCd
MSrrBBngBMgdSBMndPPMdrCBMBPnMdSPBnBCBrBnCSBPnngBCPSPBgPPMMgMgPgMCMCBdS
BSPrdgCnCrdPdSgBMrnP
```

*Bhupali:*

The same algorithm is repeated to get the evolved sequence corresponding to *Bhupali_1* and *Bhupali_2:*

```
DSSSCCCPGBCDSGSGRDSRGSRBPSDRPPSGSCRCPDBSPBPDCDCPPGSCDBSGDCSGBBCBRCSBPP
DPCCGCPBDPRBBBPCCRBBBCBGGGSSBSGCSSRCRSPPRBDGGRCBRDCGPGGCRRPGGPDGCCSCDD
GCGSCDGGCCCSRBGGPBPSCBGPSDSRCSDDGSSSPCDPSGPGGRGCRSCGGRBDPCDRPPRDCSCDSB
PCGBDRGCSDBSGGPPGPGDGSCGCDSPGDSCDGDGBPSSRSGRDPBPBGBSCDDBBGDBDRSCRPBPRP
DBSBDDCGBBCSBDCGGCPGDGPPPCRSBBCPRPCCGDPRCCCBBGDGSSSDGSCGSPGDRRSBGPDCCD
PBBCBCBBBDDCDRDCRCBPGBGPRRBGDGSGCGGBBBRSCGCGBRCGSRGCCDSGBDGBBGSSDSRPDG
BBGCBCGPRPRDGDSPCGBGDGRCGDGPBRCGDDCGDDRSGDRDDPDDGBRDBSCBPCGPDCGGRGBDCR
SPCPGDPBSSCRBSBDDCCPPRPDRDSRPRGRGGGDRSDGSCDDBBPBBCCSCGSGGCGRGCRBPCSBDB
DRDRCCDCCCGRDSCBSCSGCCBCGGGRRBSSCGBGRCRGCBGPGCBBBCDGDDBPCSPBGRDRPGSGCC
RBSSPPDDDCBSDGCDGPGPPSDRSCPDBSDCDRBPSDGPDDBPBPRPCDGPCBCPSPCDDPBPDDGCSB
RCCRDGGCRGPBSPRSRDGCCSBGRPPDSCCDDCDBDSBDDCRBPCGDCCPBBDGPCDRRPDPBDPBSBR
BPGCGPPDDDPGCRGDBDSDCRGSPBGBDBSRRCPCCRPSBDSBSGCBGDDRCRDGDGDDPSPGCCRGDR
GBSSRCSBRDBPGDBGGCBRGRPCDSPGPPBRPRBRDPRBCRRPDRDSRPGBPDBPCCGCSPBBRPCPBS
BDCSBGCSCGGDCDDCBDRDRSDDSRRBDGSSBRCGDDBGSBCBPPPPBSCBCGGRDDDSRBCGDBSGGC
SPSCPPDCSSRPRCDGDSBC
```

### *3.1 Graphical representation and Fractal Dimension:*

The graphical representation of the given two strings and the generated string are shown in the following:

***Bhairabi raga:***

The discrete sequences obtained from the music strings of Bhairabi raga consist of 9 different amplitude levels corresponding to 9 different frequencies present in the string. The strings along with the derived string are plotted below:

The two strings graph:

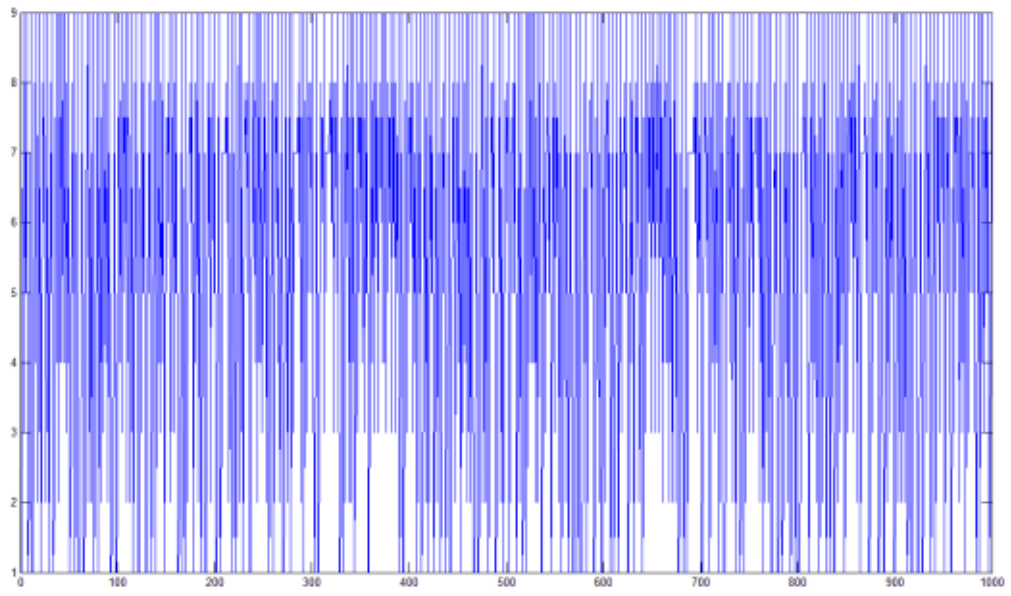

*Bhairabi_1:*

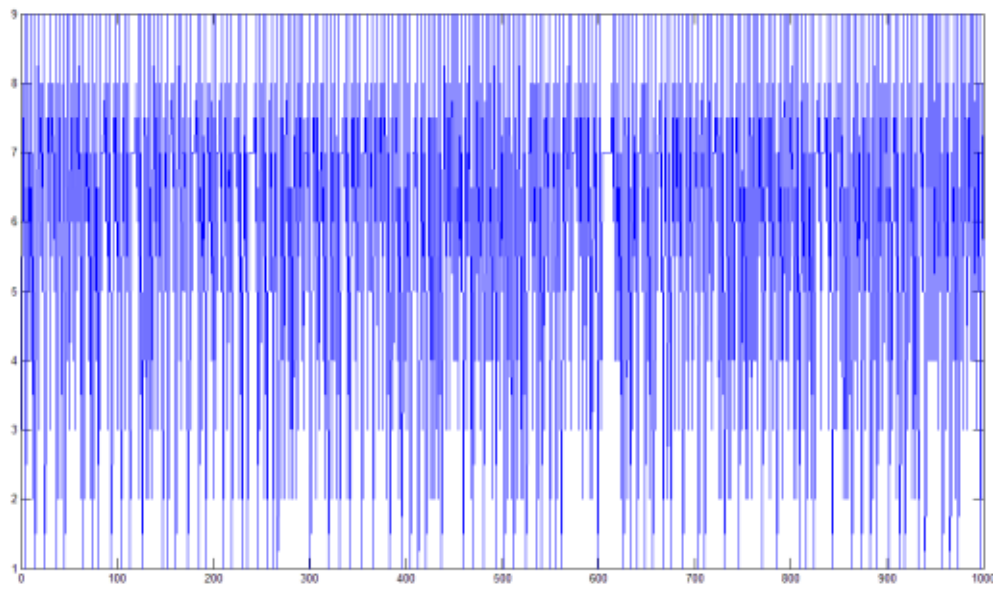

*Bhairabi_2:*

The generated string graph:

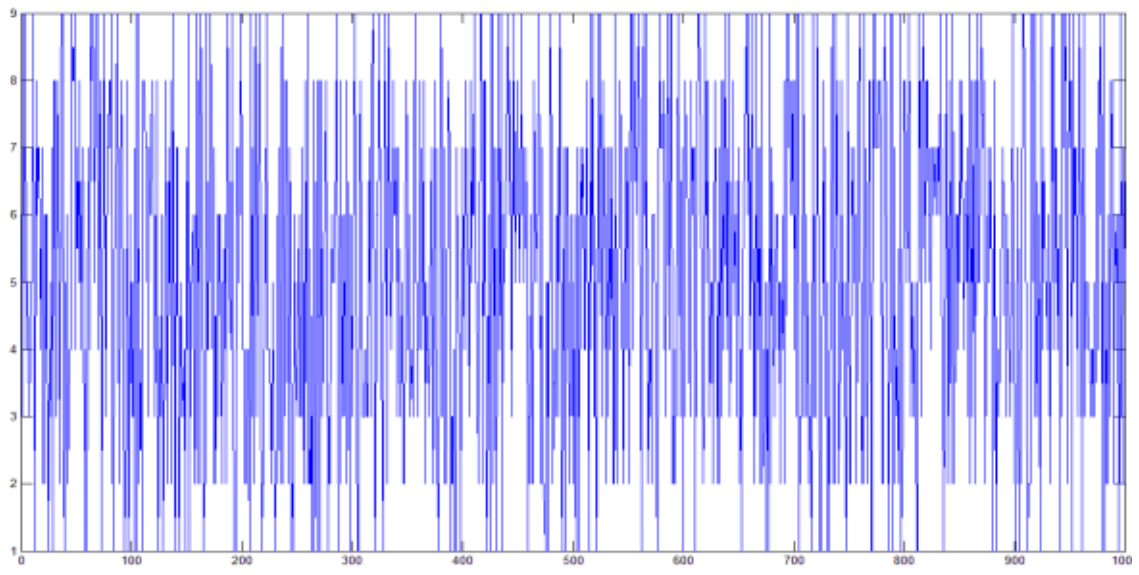

*Bhairabi_1+Bhairabi_2:*

***Bhupali raga:***

As the strings corresponding to Bhupali raga consist of7 different frequencies, the discrete sequences have 7 different amplitude levels. The strings along with the derived string are plotted below:

The string graphs are as follows:

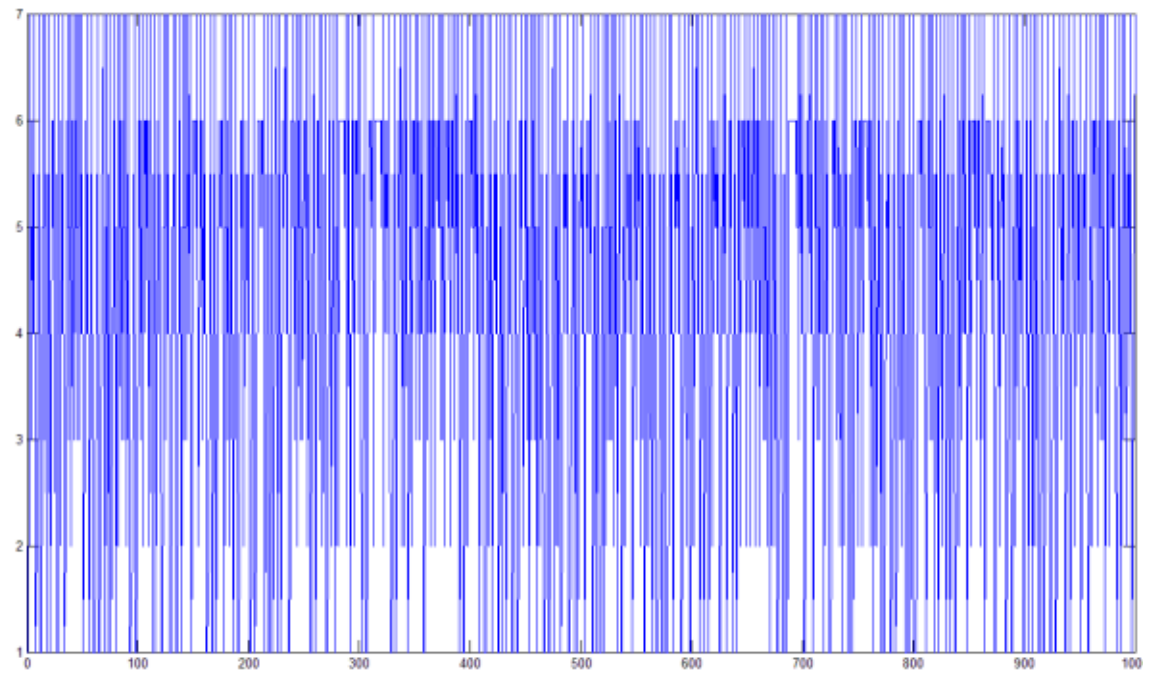

*Bhupali_1:*

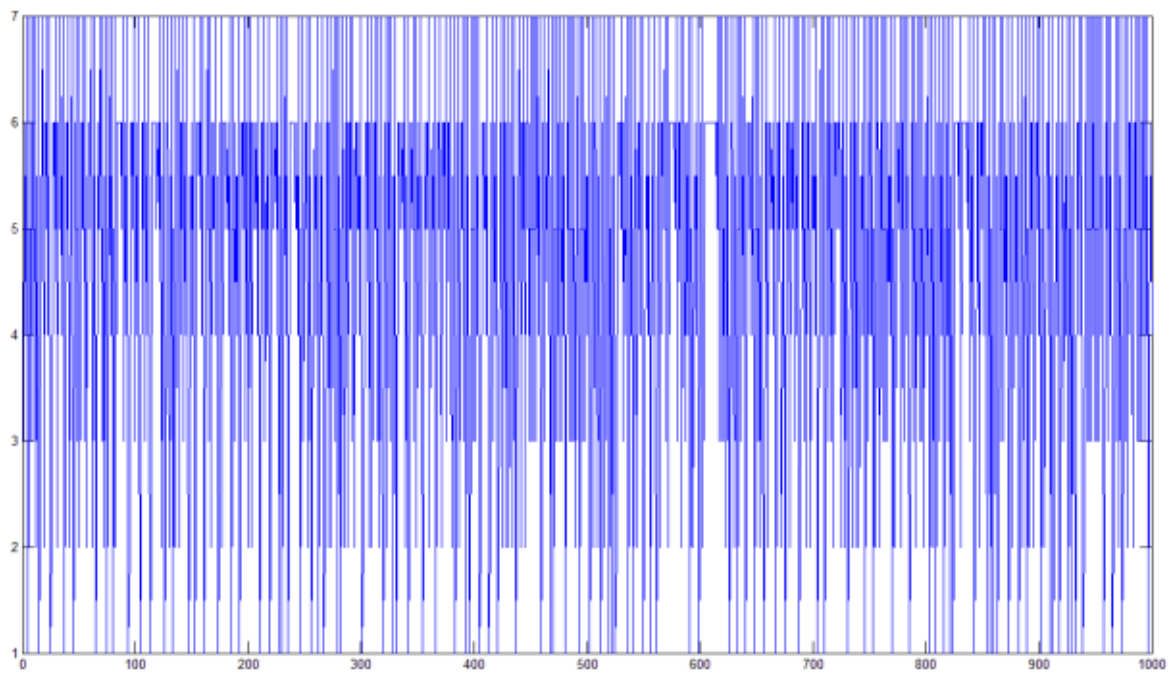

*Bhupali_2:*

The generated string:

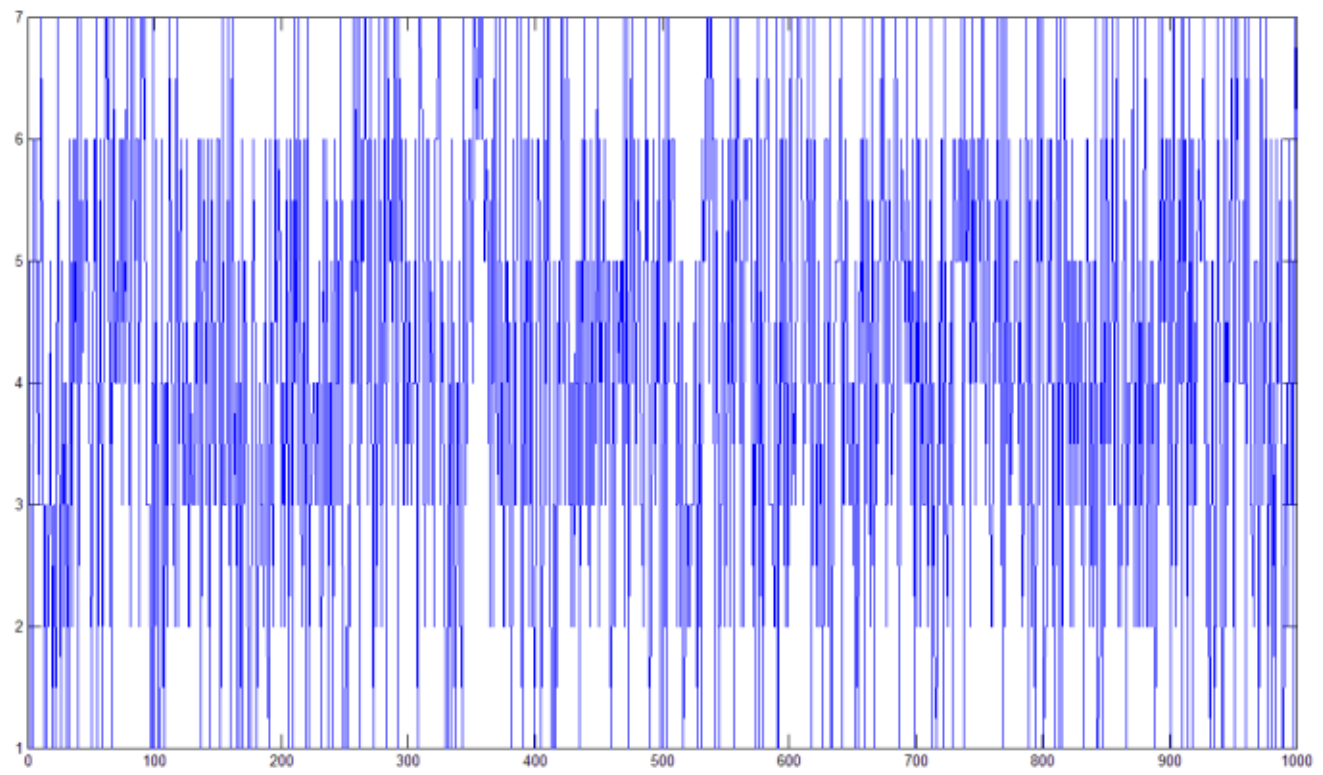

*Bhupali_1+Bhupali_2:*

The fractal dimensions of the graphs generated by logistic map (as shown above) are calculated. The fractal dimensions are 1.73752, 1.72622, and 1.79141 for the Bhupali_1, Bhupali_2, and Bhupali_1+Bhupali_2 respectively. The same for the strings corresponding to Bhairabi_1, Bhairabi_2, Bhairabi_1+Bhairabi_2 are 1.73161, 1.71914, and 1.79513 respectively. It is evident from the fractal dimension that the close relative string complexity is very close one of the twos.

## 4. *Conclusion*

In this paper, music strings are generated using logistic map utilizing its chaotic property and their analysis is presented. Statistical behavior of the generated strings is the object of further research. The search carried on in this paper is non-adaptive and in our future accomplishments, adaptive measures may be taken to ensure more closeness.

***Acknowledgement:*** The authors would like to thank to *Prof. B.S. Daya Sagar* for his valuable suggestions.

***References:***

1. Soubhik Chakraborty et al. (2008) *A Statistical Approach to Modeling Indian Classical Music Performance*, arXiv:0809.3214

2. Beth Logan and Ariel Salomon (2001) *A music similarity function based on signal analysis*, IEEE International Conference on Multimedia and Expo, pp. 952-955.ISBN 0-7695-1198-8/01 $17.00 © 2001 IEEE.

3. Web-link: http://jackhdavid.thehouseofdavid.com/papers/math.html

4. Web-link http://thinkzone.wlonk.com/Music/12Tone.htm

5. Robert M. May (1976) *Simple mathematical models with very complicated dynamics*, Nature 261, 459 – 467.

6. Little M., Heesch D. (2004). *Chaotic root-finding for a small class of polynomials*, Journal of Difference Equations and Applications 10(11): 949–953.